\documentstyle[12pt]{article}
\newcommand{\la}[1]{\label{#1}}

\newcommand{\G}{{\cal G}}
\newcommand{\cP}{{\cal P}}

\newcommand{\X}{{\cal X}}

\newcommand{\non}{\nonumber}
\newcommand{\be}{\begin{equation}}
\newcommand{\ee}{\end{equation}}
\newcommand{\ba}{\begin{eqnarray}}
\newcommand{\ea}{\end{eqnarray}}
\newcommand{\bastar}{\begin{eqnarray*}}
\newcommand{\eastar}{\end{eqnarray*}}
\newcommand{\half}{{1 \over 2}}
\begin{document}
\begin{titlepage}

\begin{flushright}
UU-ITP-15/95 \\
hep-ph/9510201
\end{flushright}

\vskip 0.7truecm

\begin{center}
{ \bf \large GAUGE VECTOR MASSES FROM FLAT CONNECTIONS?
\\
}
\end{center}

\vskip 1.5cm

\begin{center}
{\bf Antti J. Niemi $^{*\dagger}$ } \\

\vskip 0.3cm

{\it Department of Theoretical Physics,
Uppsala University \\
P.O. Box 803, S-75108, Uppsala, Sweden $^{\ddagger}$ \\

\vskip 0.2cm

{\rm and }\\

\vskip 0.2cm

Department of Physics, University of British Columbia \\
6224 Acricultural Road,
Vancouver, B.C. V6T 1Z1, Canada } \\

\end{center}

\vskip 2.0cm
\rm
\noindent
We suggest that four dimensional
massive gauge vectors could be
described by coupling ordinary Yang-Mills theory
to a topological gauge theory. For this the coupling
should excite a nontrivial degree of freedom from
the topological theory, corresponding to the
longitudinal polarization of a massive gauge vector.
If the coupling can be selected so that further degrees
of freedom are not
excited, one may entirely avoid particles such
as the Higgs. Here we discuss a simple example of this
idea, obtained by coupling standard Yang-Mills theory
to the topological gauge theory of flat connections.
We propose that our example might
describe a renormalizable
theory of massive gauge vectors with no additional
physical degrees of freedom.
\vfill

\begin{flushleft}
\rule{5.1 in}{.007 in} \\
$^{\ddagger}$  \small permanent address \\
$^{\dagger}$ \small Supported by G{\"o}ran Gustafsson
Foundation for Science and Medicine \\
\hskip 0.3cm and by NFR Grant F-AA/FU 06821-308
\\ \vskip 0.2cm
$^{*}$  \small E-mail: \scriptsize
\bf ANTTI.NIEMI@TEORFYS.UU.SE  \\
\end{flushleft}

\end{titlepage}
\vfill\eject
\baselineskip 0.65cm

The Higgs mechanism \cite{refhiggs} is the
cornerstone  in our
present understanding of mass generation.
However, even though
all other particles of the Standard Model
have been observed there is still no
experimental evidence that a Higgs particle
exists. The only indications come
from systematic theoretical constructions that have
excluded a number of alternatives in
an impressive manner {\cite{llew}:
Even though low dimensional examples of both gauge
invariant vector mass and dynamical symmetry breaking
exist, the Higgs mechanism remains the sole method for
generating a renormalizable mass for four dimensional
nonabelian gauge vectors.

In the present Letter we try to develop
an alternative to the Higgs mechanism. We suggest
that massive gauge vectors could be described by
coupling ordinary Yang-Mills theory to a topological
theory \cite{witten}, \cite{blau}. Unlike Higgs,
topological fields
do not describe physical degrees of freedom. Their
Hilbert space has only a limited number of  states.
Usually these states describe the
cohomology classes of a nilpotent
BRST operator that characterises properties of
the underlying four-manifold.
If such a theory is coupled to a conventional theory,
the coupling generically breaks the topological
invariance and nontrivial degrees of freedom are
excited. In particular,
it may happen that if a topological theory is
coupled to an ordinary Yang-Mills theory, a mass
scale is introduced and these
degrees of freedom become
the longitudinal polarization of a massive
gauge vector.  If no other degrees
of freedom are excited we may then have a
renormalizable description of massive
gauge vectors with no Higgs.

\vskip 0.3cm

Here we consider a simple example of
this idea.  We couple the standard four dimensional
$SU_Q(N)$ Yang-Mills theory with gauge
field $Q^a_\mu$ and curvature
$G^a_{\mu\nu}$ (subscripts $Q$ {\it etc.}
refer to the fields
$Q_\mu$ {\it etc.}) to
the $SU_A(N)$ $BF$ theory \cite{flats},
\cite{blau}, a topological  gauge theory that
describes flat connections
$A^a_\mu$
with curvature $F^a_{\mu\nu} \approx 0$. The $BF$
theory is particularly interesting, since it can be
viewed as a four dimensional analog \cite{blau}
of the Chern-Simons
theory that provides a gauge invariant vector
mass in three dimensions. Furthermore,
as a quantum field theory the $BF$
theory is {\it finite} \cite{flats3}.
Hence its proper coupling to ordinary Yang-Mills
theory might yield a renormalizable
quantum field theory.

In the limit where
all couplings between the two theories vanish we
have a $SU_Q(N) \times SU_A(N)$
gauge symmetry. In particular
we have two Gauss law generators corresponding to the
fields $Q_\mu$ and $A_\mu$ respectively. In this limit
the only physical degrees of
freedom are the two transverse
components  of the Yang-Mills
field $Q_\mu$, since gauge
invariance and the flatness condition eliminate all
physical excitations from the $A_\mu$ field.

We introduce a coupling between the two fields which
breaks the $SU_Q(N) \times SU_A(N)$
gauge symmetry into
the diagonal  $ SU_{Q+A}(N) \in SU_Q(N)
\times SU_A(N) $
symmetry. Consequently only
one Gauss law generator remains,
corresponding to the
diagonal $SU_{Q+A}(N)$ gauge transformations.
In the absence of a $SU_A (N)$
Gauss law constraint for the
$A_\mu$ field,  the flatness condition
is insufficient to eliminate
all of its physical excitations. The degree
of freedom that corresponds
to $SU_{Q-A} (N)$ gauge transformations
survives. This means that we are
left with three physical
degrees of freedom corresponding to
the two transverse modes of $Q_\mu + A_\mu $ and the
gauge mode of $Q_\mu - A_\mu$. If the
coupling between $Q_\mu$ and
$A_\mu$ has been selected
properly, these degrees of freedom
become the three polarizations of a massive
gauge vector. If this theory can be renormalized, we
have an alternative to the Higgs mechanism.

The four dimensional  flat connection theory
describes a $SU_A(N)$ gauge field $A_\mu^a$ subject to
the flatness condition
\be
F_{\mu\nu}^a ~=~ \partial_\mu A_\nu^a - \partial_\nu
A_\mu^a
+ f^{abc} A_\mu^b A_\nu^c ~ \approx ~ 0
\la{flatness}
\ee
and gauge invariance,
\be
{\cal G}^a ~=~ D^{ab}_{A\mu} E^b_\mu ~ = ~ \delta^{ab}
\partial_\mu E^b_\mu +
f^{acb}A^c_\mu E^b_\mu ~ \approx~ 0
\la{gausslaw}
\ee
Here we have introduced a four-dimensional (Lagrangian)
conjugate variable $E^a_\mu$ to all four components of
the gauge field $A^a_\mu$, with the (four dimensional)
Poisson bracket\footnotemark\footnotetext{We are in
a Minkowski space,
but for simplicity we do
not make a difference between upper
and lower Lorentz indices.
For example $\delta_{\mu\nu}$
in this $E , A$ Poisson
bracket is the Lorentz-covariant
${\delta^\mu}_\nu$ or more
precisely components of
a symplectic matrix, a
structure which is independent
of the metric.}
\[
\{ E^a_\mu (x) , A^b_\nu (y) \} ~\sim~ - { \delta \over
\delta A^a_\mu (x) } A^b_\nu (y) ~=~ -
\delta^{ab}_{\mu\nu} (x-y)
\]
so that (\ref{gausslaw}) indeed
generates gauge transformations of $A_\mu$
in four dimensions.
The constraints
(\ref{flatness}), (\ref{gausslaw})
obey a first class Poisson bracket algebra
\ba
\{ \  {\cal G}^a (x) , {\cal G}^b (y)  \ \}  \ & = &
\ f^{abc} \delta (x-y)
{\cal G}^c (x) \cr
\{ {\cal G}^a (x) , F^b_{\mu\nu} (y)
\ \}  \ & = & \ f^{abc}
\delta (x-y)
F^c_{\mu\nu}(x) \cr
\{ F^a_{\mu\nu} (x) ,
F^b_{\rho\sigma} (y) \}  \ & = & \ 0
\la{algebra}
\ea
However, since $F^a_{\mu\nu}$
satisfies the Bianchi identity
\be
D^{ab}_{A\mu} F^b_{\nu\rho} + D^{ab}_{A\nu}
F^b_{\rho\mu} + D^{ab}_{A\rho} F^b_{\mu\nu} =
0
\la{reducible}
\ee
the constraints (\ref{flatness}) are reducible.
Furthermore, since we also have the commutator
\be
D_{A\mu}^{ac} D_{A\nu}^{cb} -
D_{A\nu}^{ac} D_{A\mu}^{cb} ~=~
- f^{abc} F^c_{\mu\nu}
\la{2ndreducibility}
\ee
we have a first degree reducible constrained system
which is on-shell second degree reducible \cite{fv}.

We shall couple  $A_\mu^a$ to the
standard Yang-Mills
field $Q_\mu^a$ so that the coupling
introduces a mass scale, and in particular breaks
the $SU_Q(N) \times SU_A(N)$
symmetry down to the diagonal
$SU_{Q+A}(N)$ symmetry which is generated by
\be
{\cal G}^a ~=~ D^{ab}_{Q\mu}
P^b_\mu + D^{ab}_{A\mu} E^b_\mu
\la{gauss2}
\ee
Here $P^a_\mu$ is the (four dimensional) conjugate
to $Q_\mu^a$,
\[
\{ P^a_\mu (x) , Q^b_\nu (y) \} ~=~ -
\delta^{ab}_{\mu\nu} (x-y)
\]

In order to couple $Q_\mu$
and $A_\mu$ in a proper manner we
recall \cite{llew} that tree level
unitarity imposes strong
restrictions on renormalizable theories with massive
gauge vectors:
Even though Yang-Mills theory with a Proca mass is one
loop renormalizable \cite{velt}, the requirement
that tree amplitudes must be unitary
indicates that Higgs fields are almost
unavoidable \cite{llew}.

In the present case, we observe that
since both $Q_\mu$ and $A_\mu$
transform as gauge vectors
under (\ref{gauss2}) the linear combination
\be
{\cal A}^a_\mu ~=~ \frac{1}{2} ( Q^a_\mu + A^a_\mu )
\la{ourgauge}
\ee
also transforms as a gauge vector.
But since the inhomogeneous terms in the
gauge transformed
$Q_\mu$ and $A_\mu$ coincide,
the linear combination
\be
\Phi^a_\mu ~=~ Q_\mu^a - A^a_\mu
\la{ourhiggs}
\ee
transforms like a Higgs field.
If  ${\cal F}^a_{\mu\nu}$ denotes
the curvature of the gauge field ${\cal A}_\mu$,
\[
{\cal F}^a_{\mu\nu} ~=~ \partial_\mu {\cal A}_\nu^a
- \partial_\nu {\cal A}_\mu^a + f^{abc}{\cal A}_\mu^b
{\cal A}_\nu^c
\]
the {\it no-go} theorem of
\cite{llew} tells us that we should couple $Q_\mu$ and
$A_\mu$  in the following manner,
\be
S ~=~ \int \  Tr {\cal F}_{\mu\nu}^2 + \frac{1}{2}
Tr D_\mu \Phi_\nu
(D_\mu \Phi_\nu )^\dagger + \frac{m^2}{2 f^2}
Tr ({\Phi_\mu}^\dagger
\Phi_\mu - f^2 )^2 + \{ \Omega , \Psi \}
\la{action1}
\ee
Here the first term contains the standard
Yang-Mills action for $Q_\mu$, $D_\mu$
is the covariant derivative {\it w.r.t.}
(\ref{ourgauge}) and $\Omega$ is
a nilpotent BRST operator
that we shall describe
shortly: It should commute
with the action (\ref{action1})
and it should take into account both
the gauge transformations (\ref{gauss2}) and
the flatness condition (\ref{flatness}). The
functional $\Psi$ is a gauge
fermion that determines our gauge fixing.

The action (\ref{action1}) specifies our attempt to
describe
massive gauge vectors. Without the identification
(\ref{ourhiggs}) it can be viewed as a standard
renormalizable Yang-Mills-Higgs action for a $SU(N)$
gauge field ${\cal A}_\mu$ and four species of
Higgs fields $\Phi_\mu$, except that we have assigned a
negative metric to the Higgs field $\Phi_0$.
Since we try to take into account the
results of \cite{llew} as closely as possible, we
have included the $Tr \Phi^4$ self-interaction but
excluded {\it e.g.} terms like $Tr (D_\mu \Phi_\mu)^2$
which are also power-counting
renormalizable: The action
(\ref{action1}) is the most general
power-counting renormalizable action which is
consistent with a twisted version of
Lorentz transformations, with
the components of $\Phi_\mu$
transforming as scalars instead of as vectors.
Notice in particular, that we
have not (yet) included the $BF$-term.

We shall now consider the BRST operator
in (\ref{action1}). From \cite{blau}, \cite{flats},
\cite{flats2} we conclude that
it can be represented as a sum
of two nilpotent BRST operators,
$\Omega_{YM}$ describing the
conventional Yang-Mills gauge transformations and
$\Omega_{BF}$ describing the flatness condition
\be
\Omega ~=~ \Omega_{YM} ~+~ \Omega_{BF}
\la{brstsum1}
\ee
{\it i.e.} the Yang-Mills and flatness symmetries
separate in the BRST operator.

The construction of $\Omega_{YM}$
is straightforward, and
the operator $\Omega_{BF}$ has also been
discussed extensively, see {\it e.g.}
\cite{blau}, \cite{flats},
\cite{flats2}, \cite{flats3}.
Here we introduce a
slight variant of the standard
approach which is more convenient
for the present purposes.
Our construction of (\ref{brstsum1}) will be
based on the general algorithm
described in \cite{fv}, except
that we shall apply it in a Lagrangian context.
This is quite appropriate
since the constraint algebra
(\ref{algebra})-(\ref{2ndreducibility}) is manifestly
covariant. Hence  it is isomorphic
to a {\it canonical} constraint algebra in a
{\it five} dimensional
Hamiltonian theory. In particular,
the corresponding Hamiltonian BRST operator
should coincide with our four dimensional
Lagrangian BRST operator.

We first consider the BRST operator $\Omega_{YM}$
that describes the
gauge transformations
generated  by (\ref{gauss2}). We define anticommuting
ghosts $\eta^a$ and ${\cal P}^a$ with (four dimensional)
brackets
\[
\{ \eta^a (x) , {\cal P}^b(y) \} ~=~ - \delta^{ab}(x-y)
\]
We also define extra ghosts $\bar\eta^a$ and
${\bar\cP}^a$ with brackets
\[
\{ \bar\eta^a (x) , \bar\cP^b (y) \} ~=~ -
\delta^{ab}(x-y)
\]
and bosonic variables $\pi^a$, $\lambda^a$ with
\[
\{ \pi^a (x) , \lambda^b (y) \} ~=~ - \delta^{ab} (x-y)
\]
We then introduce the nilpotent
\be
\Omega_{YM} ~=~ \Omega^{min}_{YM} \
+ \ \Omega^{gf}_{YM} ~=~
\eta^a \G^a + \half f^{abc} \eta^a
\eta^b \cP^c + \lambda^a
\bar\eta^a
\la{gaugeBRST}
\ee
Here $\Omega^{min}_{YM}$ is defined by
the first two terms and describes
the algebra of (\ref{gauss2}),
while $\Omega^{gf}_{YM}$ coincides with the last term
and is necessary for gauge fixing.

We now momentarily ignore the $A_\mu$ field and
consider the standard Yang-Mills action
\be
S ~=~ \int \frac{1}{4} Tr G^2 \ + \ \{
\Omega_{YM}  ,  \Psi \}
\la{pathint}
\ee
Since $Tr G^2$ is gauge invariant, this action is
BRST invariant and
in particular the corresponding path integral is
invariant under local variations of $\Psi$. Selecting
\[
\Psi ~=~  \bar\cP^a \left( R^a (Q) + \lambda^a \right)
\]
we get
\[
S ~=~ \frac{1}{4} tr G^2 + \eta^a \{ \G^a , R^b \}
\bar\cP^b - \lambda^2 - \lambda^a R^a
\]
If we choose
\[
R^a [Q] ~=~ \sqrt{ \frac{2}{\xi} } \partial_\mu Q^a_\mu
\]
and redefine $\eta \to \sqrt{ \frac{\xi}{2} } \eta$ and
$\lambda \to \sqrt{ \frac{\xi}{2} }\lambda$ which has
unit Jacobian in the path integral, we find
by integrating over the auxiliary field $\lambda^a$
the familiar Lagrangian of  Yang-Mills theory in
the covariant $R_\xi$-gauge. This confirms that our
Lagrangian point of view works. In particular
(\ref{gaugeBRST}) with (\ref{gauss2}) is a
BRST operator that describes our $SU_{Q+A} (N)$
gauge transformations.

We now proceed to the construction of the
BRST operator that describes the flatness
constraint $F^a_{\mu\nu} \approx 0$ together with
the structure (\ref{reducible}) and
(\ref{2ndreducibility}). Following
\cite{fv} we
introduce anticommuting antisymmetric ghosts
$\psi^a_{\mu\nu}$ and $\X^a_{\mu\nu}$,
commuting ghosts $\phi^a_\mu$,
$p^a_\mu$ and anticommuting ghosts $c^a$, $b^a$.
We impose the (four dimensional)
Poisson brackets
\bastar
\{ \X^a_{\mu\nu}(x) ,
\psi^b_{\rho\sigma} (y) \} ~&=&~ - (\delta_{\mu\rho}
\delta_{\nu\sigma} -
\delta_{\mu\sigma} \delta_{\nu\rho} ) \delta^{ab} (x-y)
\non \\
\{ p^a_\mu (x) , \phi^b_\nu (y) \} ~&=&~ -
\delta_{\mu\nu}^{ab}(x-y)
\non \\
\{ \ b^a (x) , c^b(y) \} ~&=&~ -\delta^{ab} (x-y)
\eastar
and define
\[
\Omega_{BF} ~=~ \Omega_{BF}^{min} ~+~
\Omega_{BF}^{gf}
\]
\be
=~ \psi^a_{\mu\nu} G^a_{\mu\nu} +
\phi^a_\rho \epsilon_{\rho\sigma\mu\nu} D^{ab}_{Q
\sigma} \X^b_{\mu\nu} + c^a D^{ab}_{Q\mu}
p^b_\mu + \frac{1}{8} c^a f^{abc} \epsilon_{
\rho\sigma\mu\nu}
\X^b_{\rho\sigma} \X^c_{\mu\nu} ~+~
\Omega_{BF}^{gf}
\la{flatBRST}
\ee
Here $\Omega_{BF}^{min}$ describes the algebraic
structure of the
flatness condition.
The first term relates to the flatness constraint
(\ref{flatness}), the second term corresponds to
the Bianchi identity (\ref{reducible}) and the third
term takes into account the additional relation
(\ref{2ndreducibility}). But since
(\ref{2ndreducibility}) is an on-shell condition,
these three terms define
an operator which is
nilpotent only on-shell $F_{\mu\nu} \approx 0$.
The fourth term then ensures that
$\Omega_{BF}^{min}$ is {\it off-shell} {\it i.e.}
identically nilpotent.

As in (\ref{gaugeBRST}),
the (nilpotent) operator $\Omega_{BF}^{gf}$ is necessary to
fix the gauge symmetries corresponding
to the flatness condition. Reducibility
implies that besides gauge symmetries associated with
the original flatness condition we also
have additional gauge
symmetries that correspond to the following
ghost constraints
\bastar
\{ p^a_\mu  , \Omega_{BF}^{min} \} ~&=&~
\epsilon_{\mu\nu\rho\sigma}
D^{ab}_{Q\nu} \X^b_{\rho\sigma}
{}~\approx~ 0
\non \\
\{ b^a , \Omega_{BF}^{min} \} ~&=&~
- D^{ab}_{Q\mu} p^b_\mu ~\approx~ 0
\eastar
and we must define
$\Omega_{BF}^{gf}$ so that it also accounts for
these ghost constraints.
This leads to the {\it ghosts-for-ghosts} construction
\cite{fv}, which in the case of $BF$ theories
has been discussed extensively
\cite{blau}, \cite{flats}, \cite{flats2},
\cite{flats3}.

We shall not repeat this construction here.
It is (unfortunately) quite elaborate,
and will not be necessary
in the following. For us it is sufficient to know,
that the Yang-Mills symmetry and the
symmetries associated with the
flatness condition separate in the
BRST operator \cite{blau}, \cite{flats}.

We now need to combine (\ref{flatBRST})
with (\ref{gaugeBRST}). For this
we introduce the following
representations of the $SU(N)$ gauge
algebra,
\begin{eqnarray}
U^a ~ &=& ~ \half f^{abc} \X^b_{\mu\nu} \psi^c_{
\mu\nu} \non \\
V^a ~ &=& ~ f^{abc} \phi^b_\mu p^c_\mu \non \\
W^a ~ &=& ~ f^{abc} c^b b^c
\la{generators}
\end{eqnarray}
and extend $\Omega^{min}_{YM}$
in (\ref{gaugeBRST}) to
\be
\Omega_{YM}^{min} ~=~ \eta^a (
\G^a + U^a + V^a + W^a ) + \half f^{abc} \eta^a
\eta^b \cP^c
\la{2ndgaugeBRST}
\ee
This operator is nilpotent and describes the
gauge transformations of
our ghost fields. Furthermore, since
\be
\{ \Omega_{YM}^{min} , \Omega_{BF}^{min} \} = 0
\la{separate}
\ee
ensuring that the Yang-Mills and flatness symmetries
indeed separate,  we conclude that
\be
\Omega ~=~ \Omega_{YM}^{min} + \Omega_{BF}^{min}
+  \Omega_{YM}^{gf} + \Omega_{BF}^{gf}
\la{BRST}
\ee
is a nilpotent BRST operator that projects
the flatness condition
to the gauge invariant subspace. Notice in particular,
that (\ref{BRST}) leaves the
action (\ref{action1}) invariant.

We shall now proceed to fix the symmetries
in (\ref{action1}). As a consequence of the
structure (\ref{BRST}) we may proceed in
steps, by first
fixing the Yang-Mills symmetry and then the
symmetries associated with the flatness condition.

For the Yang-Mills symmetry we select
the following gauge fermion,
\be
\Psi_{YM} ~=~ \frac{1}{2}
\bar\cP^a ( \frac{1}{4}\lambda^a + \alpha \cdot
\partial_\mu A_\mu^a + \beta
\cdot \partial_\mu Q_\mu^a )
\la{Psi1}
\ee
where $\alpha$, $\beta$ specify
different gauge conditions,
and standard arguments imply that the
path integral is independent of
these parameters. For the action we get
\begin{eqnarray}
S &=& \int \  Tr {\cal F}_{\mu\nu}^2  +
\frac{1}{2} Tr
D_\mu \Phi_\nu
(D_\mu \Phi_\nu )^\dagger  +   \frac{m^2}{2 f^2}
Tr ({\Phi_\mu}^\dagger
\Phi_\mu - f^2 )^2
\non \\
&+& \ \{ \Omega_{YM} , \Psi_{YM} \} \ +
\ \{ \Omega_{BF} , \Psi_{BF} \} \non \\
&=& \int  tr {\cal F}_{\mu\nu}^2 +
\frac{1}{2} Tr D_\mu \Phi_\nu
(D_\mu \Phi_\nu )^\dagger +
\frac{m^2}{2 f^2} Tr
({\Phi_\mu}^\dagger
\Phi_\mu - f^2 )^2
\non \\
&-& \frac{\alpha + \beta}{2 \alpha} \cdot
\bar\cP^a \partial_\mu
D^{ab}_{Q\mu} \eta^b - \frac{1}{8 \alpha^2}
\lambda^2 - \frac{1}{2\alpha}\lambda^a ( \alpha
\partial_\mu A^a_\mu + \beta \partial_\mu
Q^a_{\mu} ) + \{ \Omega_{BF} , \Psi_{BF} \}
\la{S1}
\end{eqnarray}
where $\Psi_{BF}$ is a gauge fermion that fixes the
remaining flatness symmetries.

We now momentarily ignore the
$\{ \Omega_{BF} , \Psi_{BF} \}$ term
and set $\alpha = \beta = \xi$. We
then have the standard $R_\xi$-gauge
Yang-Mills-Higgs action with
four Higgs fields $\Phi_\mu$. In particular,
by eliminating $\lambda^a$ and
denoting $(A^1_\mu , A^2_\mu)
= ( Q_\mu , A_\mu )$ we find that our
gauge fields propagate according to
\be
\Delta^{ij}_{\mu\nu} ~=~ \frac{1}{2}
\left( \matrix{ 1 & 1 \cr -1 & 1 \cr } \right)
\left( \matrix{ (
\eta_{\mu\nu} - ( 1 - \frac{1}{\xi} ) { k_\mu k_\nu
\over k^2 } ) \frac{1}{k^2}  & 0 \cr 0 &
\eta_{\mu\nu} \ {1 \over k^2 - 2m^2 }
\cr } \right)
\left( \matrix{ 1 & -1 \cr 1 & 1 \cr } \right)
\la{ymhpropa1}
\ee
where we recognize the familiar
Yang-Mills and Higgs  propagators.
For the ${\bar {\cal P}} \ \eta$
ghosts we get similarly
\be
D ~=~ {2 \alpha  \over \alpha + \beta} {1 \over k^2 } \
\stackrel{ \alpha = \beta = \xi}{\longrightarrow} \
{1 \over k^2 }
\la{ghostprop}
\ee
At this point we could
redefine $\Phi^a_\mu \to \Phi^a_\mu
+ v^a_\mu$ where $v^a_\mu$ is a constant.
Selecting $v^2 = f^2$ and {\it e.g.}
improving  (\ref{Psi1})
to the 'tHooft gauge fixing condition
we then have the conventional
spontaneous symmetry breaking approach to
massive gauge fields \cite{refhiggs}.

We now return to the action (\ref{S1}),
where we may
also introduce the shift $\Phi^a_\mu
\to \Phi^a_\mu + v^a_\mu$. However, since this
corresponds to $Q_\mu^a \to Q_\mu^a +
\half v_\mu^a$ and  $A_\mu^a \to
A_\mu^a + \half v^a$, and since the last
term $\{ \Omega_{BF} , \Psi_{BF} \}$ in (\ref{S1})
depends only on $A_\mu^a$, complications will arise.
These complications may be harmless since they only
appear in BRST commutators. However,  here we prefer to
avoid them and instead we proceed by selecting
\be
\Psi_{BF} ~=~ \Psi_{BF}^{min} \ + \ \Psi_{BF}^{gf} \ = \
\frac{\gamma}{4} \
\X^a_{\mu\nu} F^a_{\mu\nu} ~+~ \Psi_{BF}^{gf}
\la{Psi2}
\ee
where $\gamma$ is another parameter, and by standard
arguments the path integral does not depend on it.
The remaining gauge fermion $\Psi_{BF}^{gf}$
denotes the ghosts-for-ghosts
contributions. For our action (\ref{Psi2}) gives
\begin{eqnarray}
S &=& \int tr {\cal F}_{\mu\nu}^2 + \frac{\gamma}{2}
tr F^2_{\mu\nu} +
\frac{1}{2} Tr D_\mu \Phi_\nu
(D_\mu \Phi_\nu )^\dagger + \frac{m^2}{2 f^2} Tr
({\Phi_\mu}^\dagger
\Phi_\mu - f^2 )^2
\non \\
&-& \frac{\alpha + \beta}{2 \alpha} \cdot
\bar\cP^a \partial_\mu
D^{ab}_{Q\mu} \eta^b + \frac{1}{2}( \alpha
\partial_\mu A^a_\mu + \beta \partial_\mu
Q^a_{\mu} )^2 + \{ \Omega_{BF}^{gf} , \Psi_{BF}^{gf} \}
\la{S2}
\end{eqnarray}
More generally (and maybe after we have extended
(\ref{2ndgaugeBRST}) so that as in (\ref{generators})
it includes an operator which gauge transforms $B_{\mu\nu}$)
we could also define
\[
\Psi_{BF}^{min} ~=~  \X^a_{\mu\nu} (
B^a_{\mu\nu} + \frac{\gamma}{4}
F^a_{\mu\nu} )
\]
where $B^a_{\mu\nu}$ is the analog of the
Lagrange multiplier $\lambda^a$
in (\ref{gaugeBRST}).
This gauge
fermion would introduce
an explicit $BF$ term in the action.
However, here we prefer
(\ref{Psi2}).

The last term in (\ref{S2}),  $\{ \Omega_{BF}^{gf} ,
\Psi_{BF}^{gf} \}$,
denotes the ghosts-for-ghosts contributions that
are necessary to fix all gauge symmetries
which are associated with the flatness condition.
This term has been analyzed extensively in the
literature \cite{blau}-\cite{flats3}, \cite{flats2}.
It introduces couplings to the gauge field $A_\mu$
and is known to have a complicated structure. However,
since the $BF$ theory is {\it finite} \cite{flats3},
this term should only yield power-counting renormalizable
couplings and renormalizable propagators. Here we
are interested in divergences that could render
(\ref{action1}) nonrenormalizable, and it is natural
to assume that such divergences should primarily
originate from
the terms that couple $A_\mu$ and $Q_\mu$. These
terms have been explicitly displayed in (\ref{S2}),
the explicit form of
$\{ \Omega_{BF}^{gf} , \Psi_{BF}^{gf} \}$
should not be relevant for the present purposes.

{}From (\ref{S2}) we get for the gauge field propagator
\be
\triangle^{ij}_{\mu\nu} ~=~  ( \eta_{\mu\nu} -
{k_\mu k_\nu \over
k^2 } )
\left( \matrix{ A & B \cr B & C \cr } \right) ~+~
{ k_\mu k_\nu \over k^2 }
\left( \matrix{ a & b \cr b & c \cr } \right)
\la{propa}
\ee
where
\ba
A ~ &=& ~ { 1 \over k^2 } \ { ( 1+
\gamma ) k^2 -  m^2
\over (1 + \gamma )  k^2 - (2 + \gamma ) m^2 }
\non \\
B ~ &=& ~ - { 1  \over k^2 } \ { m^2  \over
(1 + \gamma ) k^2 -  (2 + \gamma ) m^2 }
\non \\
C ~ &=& ~  { 1  \over k^2 } \ {k^2 - m^2 \over
(1 + \gamma ) k^2 -  (2 + \gamma ) m^2 }
\la{param1}
\ea
and
\ba
a ~ &=& ~ { 2 \over (\alpha + \beta)^2 }
\  {1 \over k^2 } \ {
( 1 + \beta^2 ) k^2 - 2 m^2 \over k^2 - 2 m^2 }
\non \\
b ~ &=& ~ { 2 \over (\alpha + \beta)^2 }
\ {1 \over k^2 } \ { (1 -
\alpha\beta ) k^2 - 2 m^2 \over k^2 - 2m^2 }
\non \\
c ~ &=& ~  { 2 \over (\alpha + \beta)^2 }
\ {1 \over k^2 } \ {( 1 + \alpha^2 ) k^2 -
2m^2 \over k^2 - 2 m^2 }
\la{param2}
\ea
We point out that (if $\gamma \not = -1 $)
for large momenta $\triangle^{ij}_{\mu\nu}$
behaves like $k^{-2}$ as it should in a
renormalizable theory. Furthermore, if we
set $\gamma \to 0$ and
$\alpha = \beta = \xi$ we get back
to (\ref{ymhpropa1}). In particular, for $\xi = 1$ we
have an analog of Feynman gauge with
the potentially troublesome $k_\mu k_\nu$ structures
in the propagator disappearing. We suggest that this
is a strong argument for renormalizability.

We now propose that the
flatness condition eliminates the Higgs field,
and (\ref{S2}) describes only a massive
vector propagator with mass $m^2$. For this
we first set $\alpha, \gamma \to \infty$ which yields
\be
\triangle^{ij}_{\mu\nu} ~=~ \left(
\matrix{ ( \eta_{\mu\nu} - { k_\mu k_\nu
\over k^2} ) \ { 1 \over  k^2 -
m^2} & 0  \cr 0  & 2 { k_\mu k_\nu \over k^2}
{1 \over k^2 - 2m^2 } \cr } \right)
\la{landau}
\ee
In this gauge the (physical) $Q_\mu$ propagates
like a mass $m^2$ gauge vector
in the Landau gauge Yang-Mills-Higgs theory, while
$A_\mu$ becomes a "gradient ghost".
Note that for an abelian theory (\ref{landau}) suggests
that our mass coincides with the
abelian Proca mass.

In the standard Yang-Mills-Higgs theory BRST invariance
ensures that the $k^2 = 0$ pole that
appears in the Landau gauge massive
vector propagator disappears.
By analogy
we then argue that this should also happen in the present
case. Furthermore, we shall now
argue that the $k^2 = 2m^2$ pole
in (\ref{landau}) must also cancel,
leaving us with the
$k^2 = m^2$ pole only. For this we
consider the $\beta,  \gamma \to \infty$ limit
of (\ref{propa}). In this limit only the $(11)$
component of $\triangle^{ij}_{\mu\nu}$ survives,
\be
\triangle^{ij}_{\mu\nu} ~=~  ( \eta_{\mu\nu} +
{k_\mu k_\nu \over
k^2 - 2 m^2} ) \ {1 \over k^2 - m^2 }
\left( \matrix{ 1 & 0 \cr 0 & 0 \cr } \right)
\la{1stlimit}
\ee
If we compare this
with the massive vector propagator in
the $R_\xi$-gauge Yang-Mills-Higgs theory
\be
\triangle_{\mu\nu} ~= ~ \left(
\eta_{\mu\nu} - (1 - \frac{1}{\xi})
{ k_\mu k_\nu \over k^2 - \frac{1}{\xi} m^2 }
\right) { 1 \over k^2 - m^2 }
\la{standprop}
\ee
we observe that (\ref{1stlimit}) corresponds
to the $\xi=\half$ gauge. In analogy with standard
Yang-Mills-Higgs theory we then argue that in our case the
$k^2 = 2 m^2$ pole must also disappear.
However, for this
we need a Ward-like identity that relates $Q_\mu$
and $A_\mu$ which is possible only
if $A_\mu$ does not entirely decouple. Indeed,
since
the $A_\mu$ self-interactions are proportional
to $\gamma$ certain diagrams containing both
$A_\mu$ and $Q_\mu$ must
survive as $\gamma \to \infty$.
These diagrams have external $Q_\mu$ lines, and are
connected to internal $A_\mu$ lines by the propagator
\be
- \frac{1}{\gamma} (\eta_{\mu\nu} - { k_\mu k_\nu
\over k^2 } ) \ \frac{1}{k^2} \ \frac{m^2}{k^2 - m^2}
\left( \matrix{ 0 & 1 \cr 1 & 0 \cr } \right) ~ + ~
{\cal O} ( { 1 \over \gamma^2 } )
\la{gamma1}
\ee
These internal $A_\mu$'s then propagate with
\be
\frac{1}{\gamma} (\eta_{\mu\nu} - { k_\mu
k_\nu \over k^2 } ) \
{ 1 \over k^2 } \left( \matrix{ 0 & 0 \cr 0
& 1 \cr } \right) ~ + ~
{\cal O} ( { 1 \over \gamma^2 } )
\la{gamma2}
\ee
and in order to produce a nontrivial $\gamma
\to\infty$ limit,
the factors of $\gamma^{-1}$ that originate from
(\ref{gamma1}), (\ref{gamma2}) must be exactly
balanced by the factors of $\gamma$ that arise from the
$A_\mu$ self-interactions
according to (\ref{S2}).
This  means that for general $\gamma$
the diagrams that contain $A_\mu$'s
and contribute to the $S$-matrix
must satisfy some Ward-like identities.
For example, if we take a derivative
of the quantum partition function {\it w.r.t.}
$\gamma$ we find the constraint (\ref{flatness}) in
the weak form  $<F^a_{\mu\nu}(x) F^a_{\mu\nu}(x)> = 0$.
In particular, the  $\gamma \to \infty$ limit is
"unitary" in the  sense that in this limit we explicitly
obtain $F^a_{\mu\nu} (x) = 0$ as a $\delta$-function
constraint in the path integral.

The previous discussion
suggests, that (\ref{S2}) is a
renormalizable action
that describes only massive
vector fields, with
no additional physical
particles: All couplings
between $Q_\mu$ and $A_\mu$
are power-counting
renormalizable
and the propagator (\ref{propa})
has the renormalizable $k^{-2}$
large momentum behavior.
Furthermore, there is also an
analog of Feynman gauge where the
potentially troublesome $k_\mu k_\nu$
structures in the gauge vector
propagators disappear. Consequently
the gauge fixing term $\{ \Omega_{BF}^{gf} ,
\Psi_{BF}^{gf} \}$ remains as the only potential
source of nonrenormalizable divergences.
This term describes the ghosts-for-ghosts
for the flatness condition, and since the $BF$
theory is {\it finite}
\cite{flats3} all interactions that emerge
from it must be power-counting
renormalizable and
all propagators must also have the
renormalizable
$k^{-2}$ behavior at large momenta.
Thus we argue that
our action (\ref{S2}) should indeed be
renormalizable. However, since
our arguments are at best
suggestive, this
needs to be confirmed either by an
explicit diagrammatic analysis or
by a general proof.

Finally we point out,
that (\ref{action1}) is not the only
possible coupling between the Yang-Mills and
$BF$ theories.
However, it appears to be the simplest one for
which all propagators behave like $k^{-2}$
for large momenta.
For example, if we only include the coupling
\[
m^2 Tr
\Phi^2 \sim m^2 Tr (A_\mu - Q_\mu )^2
\]
the propagators do
not vanish like $k^{-2}$ at large momenta.
In this sense
our construction is consistent with
the {\it no-go} theorem in \cite{llew}.
Indeed, the first three terms in
(\ref{action1}) specify the
most general action  which is invariant under a
twisted version of Lorentz transformations
where the components of $\Phi_\mu$
transform as scalars instead of as vectors.
In our final action (\ref{S2}) this twisted Lorentz
invariance is broken, but only by BRST commutators.
However, we also point out that there are
power-counting renormalizable
terms that break our twisted Lorentz
transformations and are
not BRST commutators, but
can not be directly
excluded by the arguments in
\cite{llew}.
One candidate is
$Tr (D_\mu \Phi_\mu )^2$ and
another candidate is
$Tr \Phi_\mu \Phi_\mu D_\nu \Phi_\nu $. Notice that
the latter contributes only
to the interactions.

\vskip 1.0cm

In conclusion, we have investigated if
massive gauge vectors could be described
by coupling a Yang-Mills theory to a topological
gauge theory. We have argued that if the
topological theory describes
flat connections, we get a
renormalizable theory of massive gauge vectors.
In particular, it appears that besides
the three polarizations of the
massive gauge vector there are no other physical
particles. It would be very
interesting to verify that
this conjecture is indeed correct.
Unfortunately, diagrammatic
techniques for the $BF$ theory have not yet been
developed so that an effective
perturbative investigation
would be possible. Effective diagrammatic
techniques are also
needed if we wish to investigate
the phenomenological
consequences of our proposal.

\vskip 1.0cm
We thank M. Voloshin for patiently
tutoring us in massive gauge theories.
We also thank  M. Blau, G. 'tHooft,
A. Morozov, A. Polyakov, G. Semenoff, V. Sreedhar
and L. Wijewardhana for discussions.

\end{document}